\pdfoutput=0
\documentclass[a4paper,aps,prl,twocolumn,superscriptaddress]{revtex4-2}
\usepackage[utf8]{inputenc}
\usepackage{amsmath}
\usepackage{amsfonts}
\usepackage{amssymb}
\usepackage{bm}
\usepackage{graphicx}
\usepackage{natbib}
\usepackage{units}
\usepackage{xcolor}

\begin{document}

\author{C. Arran}
\altaffiliation{Corresponding author}
\affiliation{York Plasma Institute, Department of Physics, University of York, YO10 5DD}
\author{P. Bradford}
\altaffiliation[Now at ]{University of Bordeaux}
\affiliation{York Plasma Institute, Department of Physics, University of York, YO10 5DD}
\author{A. Dearling}
\affiliation{York Plasma Institute, Department of Physics, University of York, YO10 5DD}
\author{G. S. Hicks}
\affiliation{The John Adams Institute for Accelerator Science, Blackett Laboratory, Imperial College London, London SW7 2BZ, UK}
\author{S. Al-Atabi}
\affiliation{The John Adams Institute for Accelerator Science, Blackett Laboratory, Imperial College London, London SW7 2BZ, UK}
\author{L. Antonelli}
\affiliation{York Plasma Institute, Department of Physics, University of York, YO10 5DD}
\author{O. C. Ettlinger}
\affiliation{The John Adams Institute for Accelerator Science, Blackett Laboratory, Imperial College London, London SW7 2BZ, UK}
\author{M. Khan}
\affiliation{York Plasma Institute, Department of Physics, University of York, YO10 5DD}
\author{M. P. Read}
\altaffiliation[Now at ]{First Light Fusion}
\affiliation{York Plasma Institute, Department of Physics, University of York, YO10 5DD}
\author{K. Glize}
\altaffiliation[Now at ]{Shanghai Jiao Tong University}
\affiliation{Central Laser Facility, STFC Rutherford Appleton Laboratory, Oxfordshire OX11 0QX, UK}
\author{M. Notley}
\affiliation{Central Laser Facility, STFC Rutherford Appleton Laboratory, Oxfordshire OX11 0QX, UK}
\author{C. A. Walsh}
\altaffiliation[Previously at ]{Imperial College London}
\affiliation{Lawrence Livermore National Laboratory, 7000 East Ave., Livermore, CA 94550-9234}
\author{R. J. Kingham}
\affiliation{Blackett Laboratory, Imperial College London, London SW7 2BZ, UK}
\author{Z. Najmudin}
\affiliation{The John Adams Institute for Accelerator Science, Blackett Laboratory, Imperial College London, London SW7 2BZ, UK}
\author{C. P. Ridgers}
\affiliation{York Plasma Institute, Department of Physics, University of York, YO10 5DD}
\author{N. C. Woolsey}
\affiliation{York Plasma Institute, Department of Physics, University of York, YO10 5DD}

\title{Measurement of magnetic cavitation driven by heat flow in a plasma}

\begin{abstract}
We describe the direct measurement of the expulsion of a magnetic field from a plasma driven by heat flow. Using a laser to heat a column of gas within an applied magnetic field, we isolate Nernst advection and show how it changes the field over a nanosecond timescale. Reconstruction of the magnetic field map from proton radiographs demonstrates that the field is advected by heat flow in advance of the plasma expansion with a velocity $v_N=\unit[(6\pm2)\times10^5]{m/s}$. Kinetic and extended magnetohydrodynamic simulations agree well in this regime due to the build-up of a magnetic transport barrier.
\end{abstract}

\maketitle



In extreme pressures and temperature gradients, heat flow and magnetic fields are strongly coupled, but although theoretical work shows that strong heat flows can cause significant changes in the magnetic field \cite{biermann1950, braginskii1958, haines1986}, it has long proven difficult to measure these changes experimentally. A particular challenge in magnetised plasma experiments is Nernst-driven magnetic cavitation, in which heat flow causes expulsion of the magnetic field from the hottest regions of a plasma. This reduces the effectiveness of magnetised fusion techniques \cite{sefkow2014, davies2015}, where strong magnetic fields are employed to confine the heat inside the plasma and increase yield \cite{slutz2010, hohenberger2012, perkins2017, moody2022}.

The Nernst effect is familiar in semiconductors and has been measured in semi-metals and even superconductors \cite{behnia2016}. In all of these cases, mobile charge carriers in a temperature gradient are deflected by a perpendicular magnetic field. The larger gyroradii and lower collision frequency of particles at higher temperatures results in net momentum of carriers perpendicular to both the temperature gradient and the magnetic field, establishing an electric field. In plasmas, this is typically described using classical transport theory by the thermal force acting on electrons as $\bm{\mathrm{F}_\perp} \propto - \bm{\nabla} T_e \times \bm{\mathrm{B}}$ \cite{braginskii1958}. When the Nernst electric field has a non-zero curl, the net motion of charge carriers drives advection of the magnetic field as $\partial \bm{\mathrm{B}} / \partial t = \bm{\nabla} \times (\bm{\mathrm{v_N}} \times \bm{\mathrm{B}})$, where the Nernst advection velocity is given by $\bm{\mathrm{v_N}} \approx 2 \bm{\mathrm{\phi_q}} / 5 n_e T_e$ for a heat flow $\bm{\mathrm{\phi_q}}$ \cite{nishiguchi1984, haines1986}. That is, the magnetic field is transported down temperature gradients by heat flow as well as being transported down pressure gradients by bulk plasma flow. This Nernst advection causes expulsion of the magnetic field from a hot plasma without a corresponding change in the plasma density profile, a result which cannot be explained by common models using purely ideal or resistive magnetohydrodynamics (MHD).

In general, Nernst advection is the dominant means of magnetic field transport wherever the speed of the heat flow is faster than both the bulk motion and the rate of magnetic dissipation; previous experiments which measured the Biermann battery in laser-solid interactions have shown that models of magnetised plasmas which neglect the Nernst effect fail during fast heating processes \cite{willingale2010, li2013, lancia2014, gao2015, tubman2021}. Furthermore, because heat flow depends on higher order moments of the velocity distribution, a Maxwellian approximation for heat flow is less accurate than for plasma density or current. As temperature gradients become steeper, even extended XMHD models for Nernst advection will fail. Under these non-local conditions, when the electron mean free path is no longer small compared to the length scale of the temperature gradient, neither the heat flow nor the Nernst velocity are proportional to the local electron temperature gradient. While the effect of non-locality and magnetic fields upon the temperature profile has been explored before \cite{gregori2004, froula2007,henchen2018}, non-local changes to the magnetic field have so far only been studied in kinetic simulations using Vlasov-Fokker-Planck (VFP) codes, which include the Nernst effect implicitly \cite{ridgers2008, joglekar2016}. Nernst advection therefore makes an excellent laboratory to measure kinetic effects, where changes to the heat dynamics directly affect the magnetic field.

We describe a laser-plasma experiment to measure the effect of heat flow on an applied magnetic field. Using laser-driven proton radiography \cite{borghesi2006} of an applied magnetic field, we demonstrate that Nernst advection dominates changes to the magnetic field in underdense plasmas on nanosecond timescales. Unlike previous experiments, we isolate Nernst advection and show that the magnetic field dynamics are decoupled from motion of the plasma.

\begin{figure}
\begin{center}
\includegraphics[scale=0.38,trim=0.5cm 0.5cm 0cm 3cm]{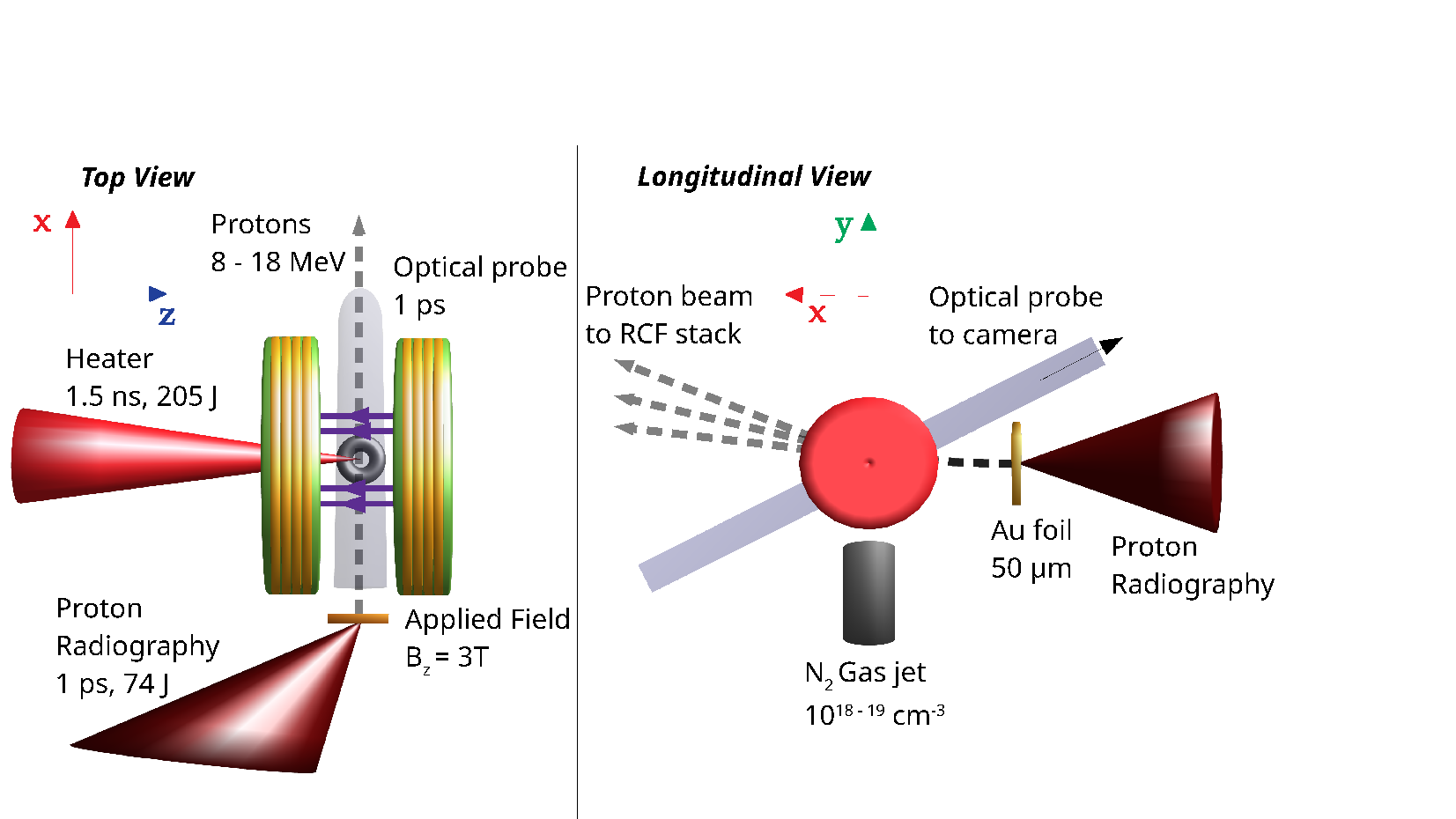}
\caption{Experimental layout shown from above (left) and from along the $z$-axis (right). The $\unit[1.5]{ns}$-duration heater beam (solid red) is focussed along the $z$-axis to a point $\unit[2]{mm}$ above the gas jet nozzle. A $\unit[1]{ps}$-duration proton radiography beam (solid red) is focussed onto a gold foil, producing a proton beam (grey dashed) which passes through the interaction point along the $x$-axis, perpendicular to the heater beam, and is deflected upwards by the applied magnetic field. The $\unit[1]{ps}$-duration collimated optical probe beam (translucent grey) also passes through the interaction point in the $x$-$y$ plane, perpendicular to the heater beam.}
\label{fig: Layout}
\end{center}
\end{figure}

We focused a $\unit[1.5]{ns}$ duration heater beam through a nitrogen gas target, propagating anti-parallel to the $\unit[3]{T}$ applied field as shown in Fig.~\ref{fig: Layout}. Laser intensities of $\unit[10^{16}]{Wcm^{-2}}$ were reached in a spot size of $\unit[19]{\mu m}$ FWHM over a Rayleigh length of $\approx\unit[1]{mm}$. This produced an approximately cylindrically-symmetric plasma with electron densities of $\unit[10^{18}-10^{19}]{cm^{-3}}$ over a scale length of $\sim\unit[100]{\mu m}$, with a temperature of around $\unit[700\pm300]{eV}$ at the highest electron density, as estimated from the measured thermoelectric field (see Supplementary Material for details). This gives a ratio between the cyclotron frequency $\omega_c$ and the collision frequency $1/\tau_e$ described by a Hall parameter around $\omega_c \tau_e \approx 1 - 10$. Under these conditions, the magnetic field and heat flow are strongly coupled, with the magnetic field restricting perpendicular heat flow, but heat flow also affecting the magnetic field dynamics. The changes in the magnetic field are described by \cite{davies2015}:

\begin{align}
\hspace{-0.5cm}\frac{\partial \mathbf{B}}{\partial t} &= \mathbf{\nabla} \times \left( \mathbf{v_B} \times \mathbf{B} \right) + {\mathbf{\nabla} \times \left( \frac{1}{\mu_0 \sigma_\perp} \mathbf{\nabla} \times \mathbf{B} \right)} + {\frac{\mathbf{\nabla} T_e \times \mathbf{\nabla} n_e}{e n_e}~},
\end{align}
where the first term describing advection is a combination of the hydrodynamic motion and the Nernst advection as $\mathbf{v_B} = \mathbf{u} - (1+\delta^c_\perp)(\mathbf{J} /e n_e) + \mathbf{v_N}$, where under our magnetised conditions the Braginskii coefficient $\delta^c_\perp\sim0.1$. This gives $\mathbf{v_B}\approx \mathbf{v_e} + \mathbf{v_N}$ \cite{walsh2020}, for electron motion $\mathbf{v_e}$. We estimate a sound speed on the scale of $\unit[10^5]{m/s}$ and a thermal diffusivity on the order of $\unit[10^4]{m^2/s}$, giving a thermal P\'eclet number of $Pe \sim 10^{-2}$. This makes heat conduction dominant over convection, indicating the importance of Nernst advection, while the Knudsen number $\lambda_\mathrm{mfp,e}/l_T \approx 1$, showing the importance of non-locality. The Braginskii conductivity is around $\sigma_\perp\sim\unit[10^7]{S/m}$, giving a magnetic Reynolds number of $Re_M\sim100$ and allowing us to neglect the magnetic diffusion and resistivity gradient flow described by the second term. The cylindrically-symmetric geometry is chosen such that the final Biermann term for generating magnetic fields is negligible, with $\mathbf{\nabla} T_e \parallel \mathbf{\nabla} n_e$. Shots without an applied magnetic field showed no magnetic field generation. Under our conditions the only possible contributions to changes in the magnetic field are therefore Nernst advection and hydrodynamic advection (`frozen-in-flow').




\begin{figure*}
\begin{center}
\includegraphics[scale=0.9,trim=0cm 0.2cm 0cm 0cm]{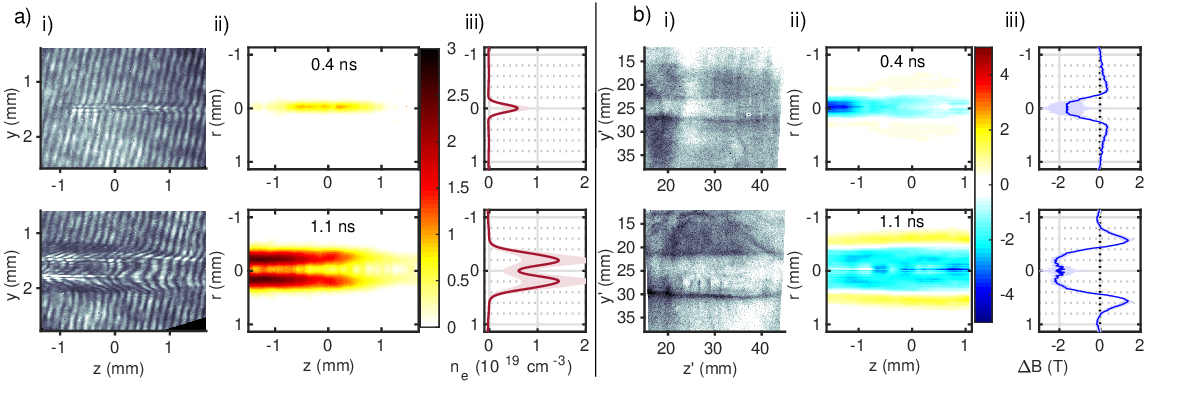}
\caption{a) Interferometry results measured $\unit[0.4]{ns}$ after the start of the heater beam (top panels) and $\unit[1.1]{ns}$ after (bottom). (i) The raw interferogram measured on the camera is shown on the left, with fringe shifts showing the presence of a long plasma column. (ii) From this interferogram we reconstruct a map of the plasma density, shown as a cross-section through the centre of the column. (iii) The longitudinally averaged mean of the radial plasma density profile is shown to the right of this, with the shaded region showing the standard deviation longitudinally. b) Radiography results measured at the same times. (i) The raw radiograph is shown on the left, with darker regions showing a higher proton dose. (ii) The change in the magnetic field reconstructed from the radiograph is shown as a cross-section through the centre of the plasma column, with (iii) the longitudinally averaged mean and standard deviation of the radial magnetic field profile to the right.}
\label{fig: Results}
\end{center}
\end{figure*}

The hydrodynamic advection was studied using optical interferometry. The $\unit[1]{ps}$-duration collimated probe beam passed transversely through the plasma column, with the interaction point re-imaged onto an Andor Neo camera after passing through a Mach-Zehnder interferometer. The interferograms (examples shown in Fig.~\ref{fig: Results}a.i) measure a plasma column much longer than it is wide, which was found to be largely symmetric by performing separate Abel inversions of the top and bottom halves of the data, before combining both halves to a symmetric map. The resulting density map and radial density profiles are shown in Fig.~\ref{fig: Results}a.ii-iii. The two laser shots were conducted under the same conditions, with the probe passing through the plasma at early and late times ($\unit[0.4]{ns}$ and $\unit[1.1]{ns}$).

The recovered density shown in the top panels of Fig.~\ref{fig: Results}a.ii-iii demonstrates that $\unit[0.4]{ns}$ after the start of the heater beam, a plasma column has been formed with a diameter of around $\unit[300]{\mu m}$ over a length of slightly under $\unit[2]{mm}$ longitudinally, with a peak density around $\unit[10^{19}]{cm^{-3}}$. At these relatively early times, the plasma is not yet fully ionized and the plasma column shows no sign of cavitation. As the heater beam continues to ionize more gas and the plasma expands, however, a density cavity forms inside the plasma column by $\unit[1.1]{ns}$ after the start of the heater beam, shown in the bottom panels.



The magnetic field evolution was measured using proton radiography performed using a broadband TNSA proton source \cite{borghesi2006}. Protons were generated by focussing a $\unit[1]{ps}$ duration laser pulse onto a $\unit[50]{\mu m}$ thick gold foil using an $f/3$ off-axis paraboloid. This proton beam passed from the foil, $\unit[20]{mm}$ from the interaction point, transversely through the plasma column, before being measured by a stack of Gafchromic EBT3 radiochromic film (RCF) $\unit[167]{mm}$ after the interaction point, giving a magnification of $9.35$. The proton intensity distribution measured by radiographs, as shown in Fig.~\ref{fig: Results}b.i, can therefore be used to reconstruct the magnetic fields through which the protons have passed \cite{kugland2012,kasim2019}. Shots taken without the applied magnetic field showed that the signal from electric fields was much weaker than the signal from magnetic fields, with proton deflections below $\unit[0.1]{mrad}$, around an order of magnitude smaller than deflections when an applied field was present.

However, the proton beam in this experiment was deflected by both the signal region within the plasma and also the constant applied magnetic field in a much larger region surrounding the plasma. This blurs out the radiographs and changes the symmetry of the signal. We therefore used a deconvolution algorithm to remove the effect of the background field, accounting for the finite energy absorption range of the RCF, the broadband proton source, and the deflection in the background field, as described in ref.~\cite{arran2021} (see Supplementary Material for more details), with a spatial resolution of around $\unit[50]{\mu m}$ for $\unit[10]{MeV}$ protons. The resulting monoenergetic radiographs were largely antisymmetric, allowing us to recover the magnetic fields separately from the thermoelectric field. The recovered change in the magnetic fields is shown in Fig.~\ref{fig: Results}b.ii-iii, where on each of the two shots the proton beam was timed such that the $\unit[10.6]{MeV}$ protons most strongly absorbed in the third layer of the RCF stack passed through the plasma simultaneously with the optical probe to within the temporal resolution of 10s of ps.

Shortly after the start of the heater beam, at $\unit[0.4]{ns}$, a strong reduction in the magnetic field strength by $\unit[-2]{T}$ in the central region is already visible in the top panel of Fig.~\ref{fig: Results}b.iii, despite no cavitation in the plasma density. The applied magnetic field is advected to the edge of the plasma by heat flow, resulting in an increase in the magnetic field strength further off-axis, at a radius of around $\unit[350]{\mu m}$. The spatial size of the cavitation is fairly uniform over a length of around $\unit[2]{mm}$, with the field cavitated over the whole of the hot plasma. This decoupling of the magnetic field profile from the plasma flow is a clear signature of the Nernst effect; this is the first time this has been measured in experiment.



\begin{figure}
\begin{center}
\includegraphics[trim=0.75cm 0.2cm 0 0]{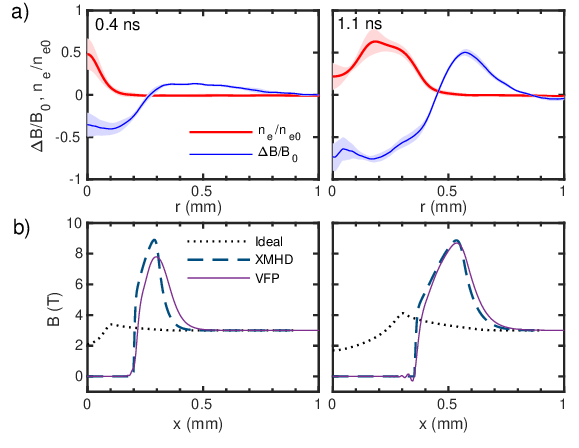}
\caption{(a) The reconstructed profiles of the magnetic field (blue) and density (red) against radius, measured at $\unit[0.4]{ns}$ (left panels) and $\unit[1.1]{ns}$ (right) and normalised to a background magnetic field $B_0 = \unit[3]{T}$ and a fully-ionized density $n_{e0} = \unit[2.4 \times 10^{19}]{cm^{-3}}$. Both profiles are longitudinally averaged over the central $\unit[1]{mm}$ of plasma, with the shaded area showing the standard deviation. (b) The magnetic field profile predicted from one-dimensional simulations, showing an ideal MHD model without the Nernst effect (dotted black line), an XMHD model including the thermoelectric Nernst term (dashed dark blue), and results from kinetic VFP simulations, with Nernst advection included implicitly (solid purple).}
\label{fig: Combined}
\end{center}
\end{figure}

The magnetic field and density profiles at these two different times are overlaid in Fig.~\ref{fig: Combined}a for comparison. The magnetic field is advected to the sheath plasma region and within the hot plasma is reduced to less than a third of its original strength. Fig.~\ref{fig: Combined}a shows that Nernst advection is significantly faster than hydrodynamic motion under these conditions. Heat flow drives cavitation in the magnetic field over a large region, before any cavitation occurs in the plasma density, with pre-heating reaching out to $r>\unit[0.5]{mm}$. We can therefore estimate the Nernst velocity at $\unit[0.4]{ns}$ by measuring the radius of the peak magnetic field at different times, reconstructed from five different proton radiographs taken on the same shot (RCF layers 2-6, absorbing proton energies from $\unit[7.6-18.3]{MeV}$). This gives a measured Nernst velocity at these early times of $\unit[(6\pm2)\times10^5]{m/s}$.

The Nernst velocity gives an estimate for the heat flux as $\phi_q = 5 n_e T_e v_N / 2$, which can be compared to the free-streaming heat flux $\phi_\mathrm{fs} = n_e T_e v_{\mathrm{th},e}$ for a thermal velocity $v_{\mathrm{th},e}$. Given that the electron thermal velocity at $\unit[700]{eV}$ is $\unit[1.6\times10^7]{m/s}$, we infer a heat flux at $\unit[0.4]{ns}$ at least one tenth of the free-streaming limit, showing the importance of correctly modelling the heat transport at these early times.  Indeed, the Braginskii estimate for the heat flow, given the measured density and temperature profiles at $\unit[0.4]{ns}$, reaches $\unit[300]{TW/m^2}$. This corresponds to a predicted Nernst velocity of $\unit[4\times10^5]{m/s}$, consistent with the measured advection.

However, the Nernst velocity falls as time goes on and the heat flow reduces, with the change in peak magnetic field position between $\unit[0.4]{ns}$ and $\unit[1.1]{ns}$ corresponding to an average advection velocity of just $\unit[(2.7\pm1.0)\times10^5]{m/s}$. Measuring the half-width at half-maximum of the density profile at $\unit[1.1]{ns}$ gives an average bulk velocity of $\approx\unit[3\times10^5]{m/s}$, which is comparable to the ion sound speed at $\unit[700]{eV}$. Whereas at early times magnetic field advection is dominated by hot electrons through the Nernst effect, at later times hydrodynamic motion on the timescale of ion motion becomes more important.


We model the plasma and magnetic field evolution with CTC \cite{bissell2010}-- an XMHD code which includes Nernst advection with a flux-limited model of heat flow -- and a kinetic VFP code, IMPACT \cite{kingham2004}, in a 1D planar geometry, to see the effects of the Nernst term and of different treatments of heat transport. Both simulations began with a uniform fully-ionized $Z=7$ plasma at a density of $n_{e0}=\unit[2.4\times10^{19}]{cm^{-3}}$ and modelled laser heating using a realistic temporal profile. Fig.~\ref{fig: Combined}b shows the predictions from IMPACT and the predictions from CTC both with and without the Nernst term. The scale of magnetic cavitation cannot be explained without invoking the Nernst effect, as ideal MHD simulations with the Nernst term turned off (shown by the black dotted line) predict only slight and slow-moving cavitation which approximately matches the density profile. Once the Nernst effect is included, however, the fluid and kinetic simulations (dashed and solid line respectively) closely agree and both capture the shape of the magnetic field profile at later times. The long tail in the magnetic field peak at $\unit[0.4]{ns}$ implies that the plasma was heated by additional processes beyond inverse bremsstrahlung over a much larger area than the initial laser spot, but as the plasma evolves the magnetic field profile shows the formation of a steeper heat front.

That the fluid and kinetic simulations agree so closely is surprising, as both simulations predict mean free paths on the order of $\unit[100]{\mu m}$, where the fluid model should break down. In the one-dimensional simulations shown here, however, the increase in the magnetic field at the edge of the hot plasma means the Hall parameter at the heat front reaches $\omega_c \tau_e \approx 10$ by $\unit[1.1]{ns}$. In this regime the heat transport becomes limited by the electron gyroradius rather than by the mean free path, with the Nernst growth rate described by Sherlock and Bissell \cite{sherlock2020} changing from ${\tilde{N} \tau_{ei} \sim (\lambda_\mathrm{mfp,e} / l_T)^2 \approx 1}$ to ${\tilde{N} \tau_{ei} \sim (r_c / l_T)^2 \lesssim 0.1}$. At early times the kinetic and fluid simulations predict different heat flows, but at later times the Nernst effect increasingly leads to a magnetic transport barrier which keeps the heat flow in a relatively local regime, even as the magnetic field inside the cavity falls to zero.



\begin{figure}
\begin{center}
\includegraphics[trim=0.7cm 0cm 0cm 0cm]{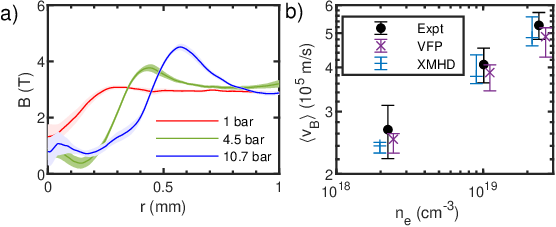}
\caption{a) The recovered magnetic field profiles measured $\unit[1.1]{ns}$ after the arrival of the heater beam for three different backing pressures, showing the mean and standard deviation of the magnetic field over a $\unit[1]{mm}$ long section in the middle of the gas jet. b) Comparing results from the simulations and the experiments, the average advection velocity of the magnetic field in the time up to $\unit[1.1]{ns}$ is measured by the position of the peak in the magnetic field.}
\label{fig: Density Scan}
\end{center}
\end{figure}

To further explore this effect, laser shots were taken at three different gas jet backing pressures. The resulting magnetic field profiles are shown in Fig.~\ref{fig: Density Scan}a, measured $\unit[1.1]{ns}$ after the start of the heater beam. At lower densities the plasma is less collisional, with a lower rate of inverse bremsstrahlung heating resulting in a colder plasma. The mean free path predicted by simulations increases to $\unit[300]{\mu m}$ at $\unit[1]{bar}$ while the maximum Hall parameter increases to $\omega_c \tau_e \approx 40$. This means both that the speed of heat flow is lower at lower densities -- leading to slower magnetic advection and less cavitation as observed in the experiment -- and also that the magnetic barrier further constrains the heat flow. In a strongly magnetised plasma we would expect the rate of heat flow to scale as $v_N \propto \tau_e^{-1} \propto n_e T_e^{-1.5}$ \cite{walsh2020}.

We estimate how the Nernst advection rates change with density by measuring the position of the peak in the magnetic field at $\unit[1.1]{ns}$ in both the experiment and 1D simulations without a flux-limiter; these results are shown in Fig.~\ref{fig: Density Scan}b. In all cases, the advection velocity falls with decreasing density, with the simulations very closely reproducing the behaviour measured in experiment. Fitting the measured average advection velocities to a power law, however, gives a trend $v_B \propto n_{e0}^{0.30\pm0.03}$. Our simulations show that $v_B \propto T_e^{0.2}$ in 1D and although the collision time is a factor of five higher at the lowest density, the advection velocity is only reduced by a factor of two. The stronger magnetisation localises the heat flow, but the Nernst advection is still faster than for a strongly magnetised plasma, particularly at early times before the magnetic barrier grows large.

In summary, we have made the first direct measurement of magnetic cavitation driven by heat flow rather than by bulk motion in the plasma. This magnetic cavitation is particularly relevant for experiments into magnetic reconnection -- where rapid heating means that magnetic transport is often Nernst-dominated -- and for inertial confinement fusion, where applied or self-generated magnetic fields have been shown to increase temperatures in the hot-spot and mitigate instability growth \cite{froula2007,chang2011,walsh2019}. As described in refs.~\cite{slutz2010, mcbride2015}, the expulsion of magnetic fields from the hottest regions of the plasma will increase the field strengths required for magnetised inertial confinement fusion techniques. We have shown that models without the Nernst term result in a spuriously high magnetic field within the plasma, and that under our moderately magnetized conditions XMHD models agree surprisingly well with kinetic simulations despite long mean free paths; the heat flow at the edge of the hot plasma remains relatively local due to the increase in the magnetic field outside the hot plasma.
\\

\begin{footnotesize}
\begin{acknowledgments}
The authors are grateful for the support of LLNL Academic Partnerships (B618488), EUROfusion Enabling Research Grants AWP17-ENR-IFE-CCFE-01 and AWP17-ENR-IFE-CEA-02, and UK EPSRC grants EP/P026796/1, EP/R029148/1, EP/M01102X/1 and EP/M011372/1. The authors particularly wish to thank the staff at Vulcan and Target Fabrication at the Central Laser Facility for all their support. Radiography calibration data from David Carroll and James Green were indispensable and we are grateful for IMPACT simulations conducted by Dominic Hill and advice with CTC from John Bissell. Part of this project was undertaken on the Viking Cluster, which is a high performance compute facility provided by the University of York. We are grateful for computational support from the University of York High Performance Computing service, Viking and the Research Computing team. The raw data and analysis code are available online at https://doi.org/10.15124/bd949fdd-1c2c-4d04-b2eb-785065198a69.

The experiment was designed by NW, CPR, and MPR and planned by CA, PB, GSH, OCE, LA, ZN, and NW. GSH and PB were the target area operators, and MN was the facility link scientist. CA led the proton radiography set-up, SA and GSH were responsible for the pulsed power magnet, OCE designed and built the optical probe layout, and KG built the system used for timing. PB, LA and MK worked on all aspects of the experiment. AD and CAW conducted fluid simulations, and AD, CA and CPR conducted VFP simulations with a code originally written by RJK. CA conducted the radiography and interferometry analysis with PB, and led the composition of the manuscript, with contributions and revisions from PB, GSH, OCE, RJK, ZN, CPR and NW.
\end{acknowledgments}
\end{footnotesize}

%

\end{document}